\documentclass[10pt,aps,twocolumn,nofootinbib]{revtex4-1}
\usepackage{amsfonts}
\usepackage{mathrsfs}
\usepackage{amsmath}
\usepackage{amssymb}
\usepackage[normalem]{ulem}
\usepackage{extarrows}
\usepackage{slashed}
\usepackage{isodateo}
\usepackage{graphicx} 
\usepackage{xcolor}
\usepackage[bookmarksnumbered=true,bookmarksopen=true]{hyperref}
 \hypersetup{colorlinks,%
             linkcolor=[rgb]{0,0.3,0.6}, %
             citecolor=[rgb]{0,0.3,0.6}, %
             urlcolor=[rgb]{0,0.3,0.6}}%
\renewcommand{\rm}{\mathrm} 
\begin{document}

\title{Probing Unification With Precision Higgs Physics}
\author{Sibo Zheng$^{1,2,}$}
\email{Email: sibozheng.zju@gmail.com}
\affiliation{$^{1}$Department of Physics, Chongqing University, Chongqing 401331, China \\
$^{2}$Department of Physics, Harvard University, Cambridge, MA 02138, USA}

\begin{abstract}
We propose a novel approach of probing grand unification through precise measurements on the Higgs Yukawa couplings at the LHC.
This idea is well motivated by the appearance of effective operators not suppressed by the mass scale of unification $M_{\rm{U}}$ in realistic models of unification with the minimal structure of Yukawa sector.
Such operators modify the Higgs Yukawa couplings in correlated patterns at scale $M_{\rm{U}}$ that hold up to higher-order corrections.
The coherences reveal a feature that, the deviation of tau Yukawa coupling relative to its standard model value at the weak scale 
is the largest one among the third-generation Yukawa couplings.
This feature, if verified by the future LHC, can serve as a hint of unification.
\end{abstract}
\maketitle

\section{Introduction}
The Higgs Yukawa couplings to the standard model (SM) fermions such as top \cite{1804.02610}, bottom \cite{1709.07497}, and tau \cite{1501.04943}
have been verified at the LHC. 
With upcoming data at the future LHC \cite{1310.8361} or ILC \cite{Asner:2013psa} and CEPC-SPPC \cite{CEPC} in preparation, 
we will enter into an era of precision measurements on the Higgs,
which may shed light on the fundamental laws underlying the electroweak symmetry breaking.
While the minimal version of supersymmetry is not satisfactory such as in explaining the observed Higgs mass,
it is still on the short list of frameworks that address some well-known big questions such as the hierarchy problem.

Although supersymmetry is advocated to solve the hierarchy problem, 
there are limited tools to probe the underlying grand unification theory (GUT) behind supersymmetry,
as the unification scale $M_{\text{U}}$ is far larger than the weak scale.
Until now, proton decay is the most important tool to directly detect unification 
(For a review on unification and proton decay, see ref.\cite{review}.).
Nevertheless, 
 a large amount of models easily evade the Super-Kamiokande limits \cite{0903.0676,9904020} on the proton decay by adjusting the value of $M_{\text{U}}$ in the mass range $10^{15-17}$ GeV according to the dependence of proton decay lifetime on $M_{\text{U}}$, i.e.
 $\tau_{p}\sim 1/M^{4}_{\text{U}}$.
A delay of update on those experimental limits has postponed our exploration along this line.

In this work, we propose a novel method of probing GUT through precision measurements on the Higgs Yukawa couplings.
Similar to high-dimensional operators that lead to proton decay through interactions between the GUT-scale states and the SM fermions,
there are analogies which modify the Higgs Yukawa couplings due to interactions between GUT-scale states $\phi$ and the Higgs doublets: 
\begin{eqnarray}{\label{eoperator}}
W_{\text{eff}} &\simeq& \int d^{2}\theta \left[(y^{i}_{u}+\epsilon^{i}_{u})Q_{i}\bar{u}_{i}H_{u}+(y^{i}_{d}+\epsilon^{i}_{d}) Q_{i}\bar{d}_{i}H_{d}\right.\nonumber\\
&+&\left. (y^{i}_{e}+\epsilon^{i}_{e})L_{i}\bar{e}_{i}H_{d}+\cdots\right],
\end{eqnarray}
where $i$ is the generation index, 
$y_{f}$ refer to the SM values of Yukawa couplings at the weak scale with $f=\{u,d,e\}$,
and the corrections at the weak scale are denoted by $\epsilon^{i}$.
What we have neglected in Eq.(\ref{eoperator}) are higher-order effective operators suppressed by power laws of $1/M_{\text{U}}$.
We will show that
$a)$. $\epsilon^{i}$ terms are less than unity but not suppressed by $1/M_{\text{U}}$ for
\begin{eqnarray}{\label{smallp}}
\epsilon^{i}\sim \frac{\langle \phi\rangle}{M_{\text{U}}},
\end{eqnarray}
with $\langle \phi\rangle\sim M_{\text{U}}$ referring to the vacuum expectation value (vev) of SM singlet responsible for breaking the GUT gauge group.
$b)$ All of $\epsilon^{i}$ terms are always correlated in specific patterns rather than independent parameters.
$c)$. The coherences lead to specific patterns of derivations in the Higgs Yukawa couplings from their SM reference values,
which may be verified by the future LHC in certain parameter ranges.

\begin{figure*}
\centering
\includegraphics[width=14cm,height=4cm]{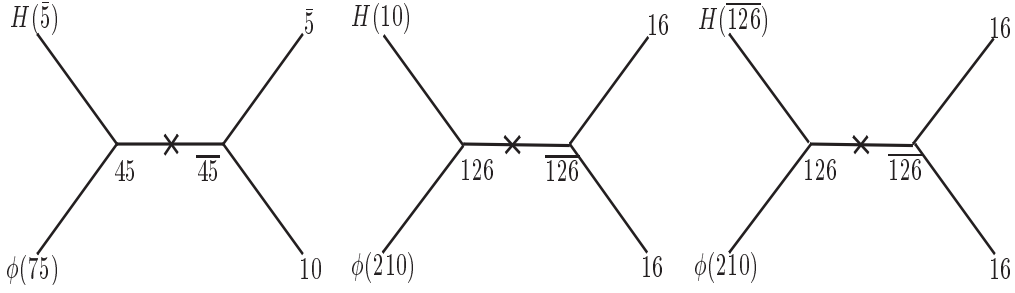}
\caption{Supergraphs for effective operators unsuppressed by the scale $M_{\text{U}}$ in Eq.(\ref{eoperator}) in the case of $\rm{SU}(5)$ (\emph{left}) 
and $\rm{SO}(10)$ (\emph{middle, right}) with the minimal Yukawa sector, where crossings refer to the effective vectorlike mass insertions.
In each diagram, inserting $\phi$ with a number of $n$ into the internal  line of propagator yields high-order contributions multiplied by $\epsilon^{n}$. }
\label{feyn}
\end{figure*}

\section{Minimal Yukawa Sector}
Let us begin our study with a briefly review on the realistic models of unification with the minimal Yukawa sectors.
For the $\rm{SU}(5)$ unification, 
the Higgs fields in the minimal Yukawa sector are composed of a $5$,  a $\bar{5}$ and a $\overline{45}$.
The $\overline{45}$ \cite{GJ} is added to the Yukawa sector in order to adjust the lepton and down quark Yukawa couplings at the scale $M_{\text{U}}$.
Under this Yukawa structure, the Yukawa couplings at scale $M_{\text{U}}$ are given by,
\begin{eqnarray}{\label{benchmark1}}
y^{i}_{u}&=&\frac{\upsilon^{5}_{u}}{\upsilon_{u}}Y_{u}^{ij},\nonumber\\
y^{i}_{d}&=&\frac{\upsilon^{\bar{5}}_{d}}{\upsilon_{d}}Y_{d}^{ij}+\frac{\upsilon^{\overline{45}}_{d}}{\upsilon_{d}}Y^{ij}_{45}, \nonumber\\
y^{i}_{e}&=&\frac{\upsilon^{\bar{5}}_{d}}{\upsilon_{d}}Y_{d}^{ij}-3\frac{\upsilon^{\overline{45}}_{d}}{\upsilon_{d}}Y^{ij}_{45},
\end{eqnarray}
where $\upsilon^{5}_{u}$, $\upsilon^{\bar{5}}_{d}$ and $\upsilon^{\overline{45}}_{d}$ refers to the vevs of doublets in $5$, $\bar{5}$ and $\overline{45}$, respectively;
$\upsilon_{u}=\upsilon\sin\beta$, $\upsilon_{d}=\upsilon\cos\beta$, with $\upsilon= 174$ GeV;
and $Y_{u}$, $Y_d$ and $Y_{45}$ are 3$\times$ 3 matrixes in generation space, with $i,j=1-3$.
We recall that with $m_{45}\sim M_{\text{U}}$, 
the proton decay mediated by the component fields in $\overline{45}$ is small.
What matters \cite{1809.08724} in this Yukawa sector is the generation of a small vev $\upsilon^{\overline{45}}_{d}$, compared to a  large mass $m_{45}$.

For the $\rm{SO}(10)$ unification, the minimal Yukawa sector \cite{1805.10631}  is composed of a $10$ and an $\overline{126}$.
The purpose of $\overline{126}$ \cite{9209215} closely follows from that of $45$ in the $\rm{SU}(5)$.
The Yukawa couplings at scale $M_{\text{U}}$ read as,
\begin{eqnarray}{\label{benchmark2}}
y^{i}_{u}&=&\frac{\upsilon^{10}_{u}}{\upsilon_{u}}Y_{10}^{ij}+\frac{\upsilon^{\overline{126}}_{u}}{\upsilon_{u}}Y^{ij}_{126},~~~~
y^{i}_{d}=\frac{\upsilon^{10}_{d}}{\upsilon_{d}}Y_{10}^{ij}+\frac{\upsilon^{\overline{126}}_{d}}{\upsilon_{d}}Y^{ij}_{126}, \nonumber\\
y^{i}_{e}&=&\frac{\upsilon^{10}_{d}}{\upsilon_{d}}Y_{10}^{ij}-3\frac{\upsilon^{\overline{126}}_{d}}{\upsilon_{d}}Y^{ij}_{126},~~
y^{i}_{\nu}=\frac{\upsilon^{10}_{u}}{\upsilon_{u}}Y_{10}^{ij}-3\frac{\upsilon^{\overline{126}}_{u}}{\upsilon_{u}}Y^{ij}_{126},\nonumber\\
\end{eqnarray}
where $\upsilon^{10}_{u,d}$ and $\upsilon^{\overline{126}}_{u,d}$ refers to the vevs of doublets in $10$ and $\overline{126}$, respectively, and $y_{\nu}$ is the neutrino Yukawa coupling.
Similar to $\overline{45}$, for $m_{126}\sim M_{\rm{U}}$ the proton decay due to component fields of $\overline{126}$ is small.

Fitting the values of Yukawa couplings at scale $ M_{\rm{U}}$ 
to their SM values at the scale $m_Z$ in terms of the renormalization group equations (RGEs),
one can fix all of the input parameters in Eq.(\ref{benchmark1}) or Eq.(\ref{benchmark2}).

\section{Unsuppressed Effective Operators}
It is well known that we will obtain the effective operators \cite{0301121,0707.0005} which contribute to Eq.(\ref{eoperator}) 
after one integrates out heavy freedoms with characteristic mass scale $M$.
One is also aware of that the ability of testing those effective operators dramatically declines as the value of $M$ increases. 
In the situation $M\sim M_{\text{U}}$, the effects on SM observables due to those operators are supposed to be tiny (e.g., $\tau_p$), unless they are not suppressed by power laws of $1/M_{\text{U}}$.
We will show that there are indeed unsuppressed effective operators in Eq.(\ref{eoperator}) by integrating out the heavy Higgs field $45$ or $126$ in the minimal Yukawa structure as discussed in the preceding section.

We show in the \emph{left} plot of Fig.\ref{feyn} the generation of the effective operator
\begin{eqnarray}{\label{operator1}}
(\text{I}): \delta W_{\text{eff}}&\sim&\int d^{2}\theta \frac{Y^{ij}_{45}}{m_{45}}\phi(75) \bar{H}(\bar{5}) \psi_{i}(\bar{5})\psi_{j}(10),
\end{eqnarray}
after integrating out the vectorlike Higgs fields $45$ in the minimal Yukawa sector\footnote{In this minimal Yukawa sector, fine tuning is required in order to keep the Higgs doublets light.}.
Here, a SM singlet with nonzero vev in the $75$-dimensional Higgs plays the role of $\phi$ in Eq.(\ref{smallp}). 
The nonzero vev that can be obtained e.g, through a self-interaction in the GUT-scale superpotential  $W\supset \phi^{3}(75)$ spontaneously breaks the $\rm{SU}(5)$ gauge group into the SM gauge group.
Afterwards, it gives rise to the effective operator in Eq.(\ref{operator1}) through the interaction $W\supset \phi(75)H(\bar{5})H(45)$
consistent with the $\rm{SU}(5)$.

Substituting $\phi(75)$ with vev $\langle\phi(75)\rangle$ in Eq.(\ref{operator1}), one finds the coefficients in Eq.(\ref{eoperator}) 
\begin{eqnarray}{\label{coherence1}}
(\text{I})&:& \epsilon^{ij}_{d} \simeq \epsilon^{ij}_{e} \simeq \epsilon Y^{ij}_{45}\frac{\upsilon_{d}^{\bar{5}}}{\upsilon_{d}},
\end{eqnarray}
where the overall scale $\epsilon=\langle\phi(75)\rangle/M_{\text{U}}$, with $M_{\text{U}}$ referring to the effective VL mass $m_{45}$. 
For simplicity, all Yukawa coefficients in the vertexes of the Feynman diagram are absorbed into $\epsilon$.
We observe an important coherence $\epsilon^{ij}_{d} \simeq \epsilon^{ij}_{e} $ in Eq.(\ref{coherence1}),
which is a result of the $\rm{SU}(5)$ embedding and independent of GUT-scale parameters such as $\epsilon$, $Y_{45}$ and the ratio of two vevs.
This coherence can be a key to reveal the $\rm{SU}(5)$ unification through the precision measurements on relevant Higgs Yukawa couplings.

Similarly, we can analysis the Feynman diagrams in the \emph{middle} and \emph{right} plots of Fig.\ref{feyn} respective to the $\rm{SO}(10)$,
\begin{eqnarray}{\label{operator2}}
(\text{II}): \delta W_{\text{eff}}&\sim&\int d^{2}\theta \frac{Y^{ij}_{126}}{m_{126}} \phi(210) H(10) \psi_{i}(16)\psi_{j}(16), \nonumber\\
(\text{III}): \delta W_{\text{eff}}&\sim&\int d^{2}\theta \frac{Y^{ij}_{126}}{m_{126}}  \phi(210) \bar{H}(\overline{126}) \psi_{i}(16)\psi_{j}(16),\nonumber\\
\end{eqnarray}
where the vectorlike Higgs is $126$ instead of $45$, 
$\phi$ is $210$ instead of $75$,
and a $16$ supermultiplet represents a whole generation of SM fermions.
Similarly to $75$ in the $\rm{SU}(5)$, the singlet vev of $210$ in the $\rm{SO}(10)$ can be obtained through self-interaction $W\supset \phi^{3}(210)$. 
The type-II and type-III effective operator in Eq.(\ref{operator2}) are then mediated by the interaction $W\supset \phi(210)H(10)H(126)$ and $W\supset \phi(210)H(126)H(\overline{126})$ consistent with the  $\rm{SO}(10)$, respectively. 
 
Unlike in the $\rm{SU}(5)$, in this minimal Yukawa sector the light Higgs doublets can be dynamically obtained as in the benchmark model studied below.
From Eq.(\ref{operator2}), we have
\begin{eqnarray}{\label{coherence2}}
(\text{II}):\epsilon^{i}_{u}&\simeq&\epsilon^{i}_{\nu}=\epsilon Y^{i}_{126}\frac{\upsilon_{u}^{10}}{\upsilon_{u}},~~~~
\epsilon^{i}_{d}\simeq\epsilon^{i}_{e} \simeq\epsilon Y^{i}_{126}\frac{\upsilon_{d}^{10}}{\upsilon_{d}};\nonumber\\
(\text{III}):\epsilon^{i}_{u}&\simeq&-\frac{\epsilon^{i}_{\nu}}{3}=\epsilon Y^{i}_{126}\frac{\upsilon_{u}^{\overline{126}}}{\upsilon_{u}},~~
\epsilon^{i}_{d}\simeq-\frac{\epsilon^{i}_{e}}{3}\simeq\epsilon Y^{i}_{126}\frac{\upsilon_{d}^{\overline{126}}}{\upsilon_{d}},\nonumber\\
\end{eqnarray}
where $\epsilon=\langle\phi(210)\rangle/M_{\text{U}}$, with $M_{\text{U}}$ referring to the effective VL mass $m_{126}$.
Similar to the the coherence in Eq.(\ref{coherence1}), we observe two new types of coherences in Eq.(\ref{coherence2}) that 
result from the $\rm{SO}(10)$ embedding. 

One may ask how general the coherence(s) can be in the two classes of models with the minimal Yukawa sector. 
The key factor is the SM singlets within high dimensional representations of GUT group behind the sector which controls the behavior of GUT breaking. 
If the singlet vev in $75$- ($210$-) dimensional Higgs uniquely breaks the $\rm{SU}(5)$ ($\rm{SO}(10)$),   
the type-I (II-III) coherence(s) will hold up to higher-order corrections, as illustrated in Fig.\ref{feyn}.
Example in this situation include the well known the $75$-dimensional Higgs used as an economic solution to the doublet-triplet problem \cite{0007254}. 
Moreover, we expect the coherence(s) still valid even in the case where multiple singlet vevs collectively break the GUT group 
but compared to $75$ or $210$ the others are subdominant. 
Examples in this case include a benchmark model discussed below. 
Finally, the validity of these coherences may be likely even for a subdominant singlet vev of $75$ or $210$ when the  analogies similar to the effective operators in Fig.\ref{feyn} due to the dominant vevs are produced at higher-loop orders.

\section{Precision Measurements}
Either in Eq.(\ref{coherence1}) or Eq.(\ref{coherence2}), 
the corrections to Higgs Yukawa couplings dominate the next-leading order contributions,
as long as the Yukawa sector is minimal and $\epsilon$ is less than unity.
Their impacts at the scale $m_Z$
rely on the magnitudes of orders of these corrections.
In individual situation, there are small hierarchies among the matrix elements of $Y^{ij}_{45}$ or $Y^{ij}_{126}$ \cite{1805.10631}, which arise from the SM fermion mass hierarchies.
As a result, the largest effects always occur in the third-generation Yukawa couplings $y_{\alpha}$ ($\alpha=t,b,\tau$).
Since precision measurements on Yukawa couplings $y_{\alpha}$ are prior to the first two generations at the LHC,
we will mainly focus on $y_{\alpha}$ as what follows. 

Given an explicit $\epsilon$, the weak-scale effects on $y_{\alpha}$ can be derived as follows.
First, one uses the central values \cite{1306.6879} of SM observables at the scale $m_Z$
to determine all input parameters at the scale $M_{\rm{U}}$ through the RGEs from $m_Z$ to $M_{\text{U}}$. 
During this process, one has to deal with various intermediate effective theories.
Second, we add correlated $\epsilon$-corrections to $y_{\alpha}$ at the scale $M_{\rm{U}}$,
then perform the RGEs reversely from the scale $M_{\rm{U}}$ to $m_Z$, 
which gives rise to the dependences of $\delta y_{\alpha}$ on $\epsilon$ at the scale $m_Z$.
During the RGEs, there are certain uncertainties similar to proton decay\footnote{Compared to proton decay, 
the theoretical uncertainties in our approach are theoretically improved in the sense 
that the dimensionless parameters $\delta y_{\alpha}$ are logarithm- rather than power-law dependent on mass scales such as $M_{\rm{U}}$.}.

\begin{figure*}
\centering
\includegraphics[width=8cm,height=8cm]{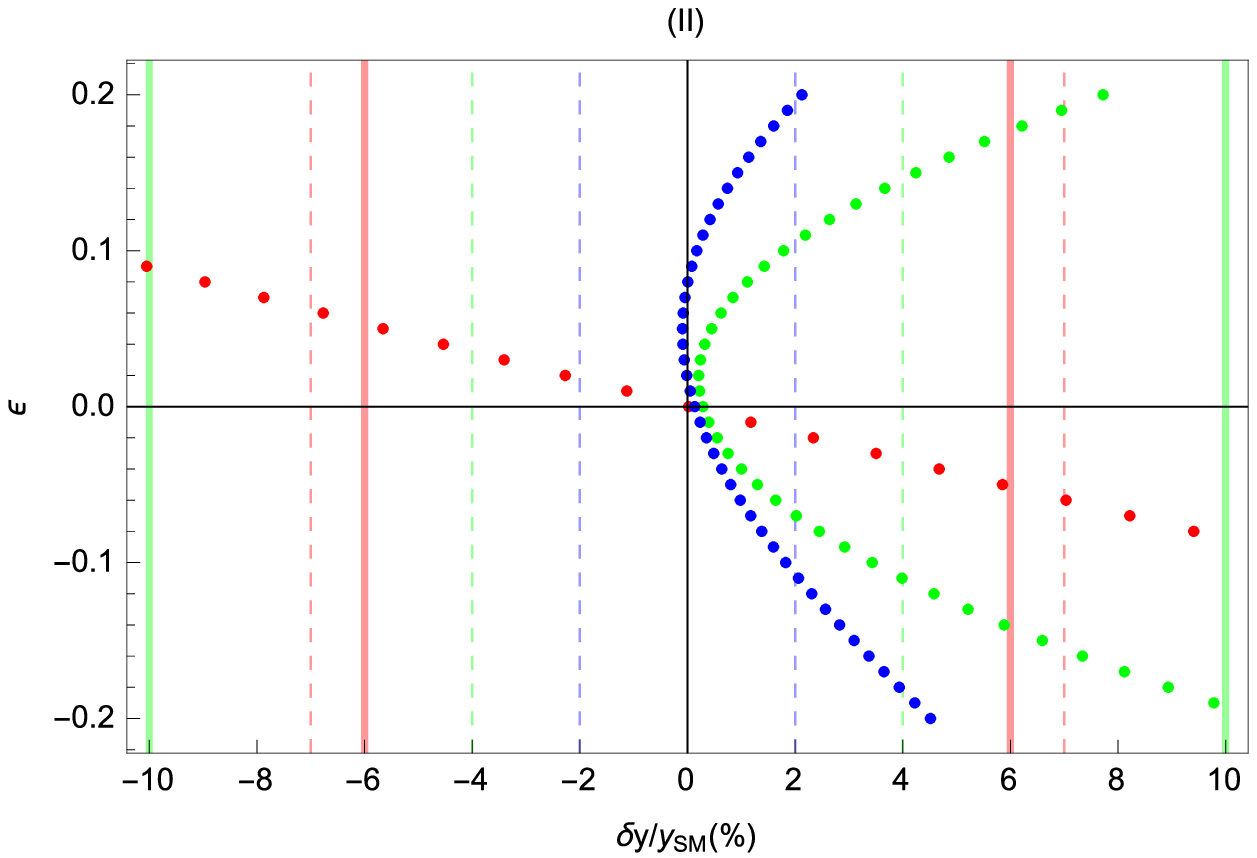}~~~~~~
\includegraphics[width=8cm,height=8cm]{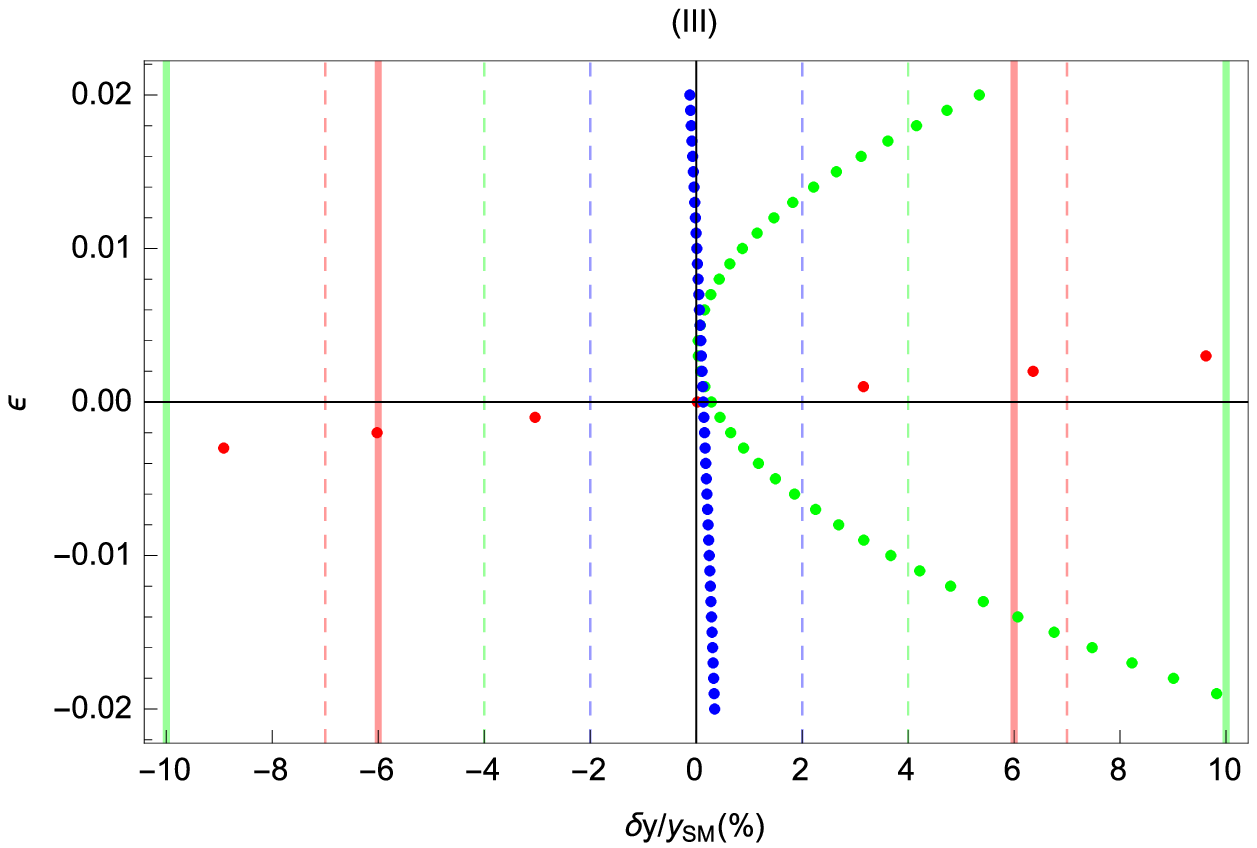}
\caption{Deviations of Yukawa couplings $y_{t}$ (dotted blue), $y_{b}$ (dotted green) and $y_{\tau}$ (dotted red) relative to their SM expectation values in the case of type-II corrections $\epsilon=\epsilon_{t}/r=\epsilon_{b}=\epsilon_{\tau}$ (\emph{left}) and the type-III corrections 
$\epsilon=-\epsilon_{t}/(r\cdot s)=\epsilon_{b}=-\epsilon_{\tau}/3$ (\emph{right}), respectively. 
The vertical thick and dashed lines refer to LHC limits \cite{1310.8361} with luminosity 300 fb$^{-1}$ and 3000 fb$^{-1}$, respectively, where the references of colors are the same as the points. 
We have chosen the SM expectation values at  the scale $m_Z$ in ref.\cite{1306.6879},  $M_{\rm{U}}=2\times10^{16}$ GeV, and $\tan\beta=10$. 
Deviations larger than $10\%$ from the SM expectation values are not shown, some of which have been already  disfavored by the LHC data \cite{1804.02610,1709.07497,1501.04943}.
See text for details.}
\label{measure}
\end{figure*}

For the $\epsilon$ terms in explicit models,
there is a lack of ``complete" fit to the $\rm{SU}(5)$ model with the minimal Yukawa sector in the literature.
The ``completeness'' means that all SM fermion masses and mixings are addressed, 
with important constraints such as proton decay taken into account.
Otherwise, the theoretical uncertainty is too large to invalidate the RGE analysis.
For earlier studies on this model, see. e.g. refs.\cite{9809554,0702287}.
On the contrary, there are extensive studies on the $\rm{SO}(10)$ model with the minimal Yukawa sector.
We will use the latest results in ref.\cite{1805.10631}, while earlier studies can be found, e.g. in refs.\cite{so1, 0405300,1102.5148, so2}.

The benchmark $\rm{SO}(10)$ model is composed of following effective theories for various RG scales $\mu$: 
$\rm{i})$ the SM for $\mu<m_{Z}$; 
$\rm{ii})$ SM with gauginos for $m_{Z}<\mu <1$ TeV;
$\rm{iii})$ the split-supersymmetry for $1$ TeV $<\mu<10^{2}$ TeV;
$\rm{iv})$ the complete MSSM for $10^{2}$ TeV $<\mu<m_{\nu_{R}}$, 
where $m_{\nu_{R}}$ denotes the right-hand neutrino mass threshold;
and finally $\rm{v})$ the MSSM with extended gauge group $U(1)_{B-L}$ for $m_{\nu_{R}} <\mu < M_{\text{U}}$.
Compared to the $\rm{SU}(5)$,  there are more intermediate RGE steps in this model.
Here, a few comments are in order. 
First,  for simplicity $M_{\text{U}}$ that relies on the scalar vevs of $210$ (together with $54$) is fixed to be $2\times 10^{16}$ GeV.
Second, the right-hand neutrino mass $m_{\nu_{R}}$ which is determined by the vev of $126$ \cite{0405300} is in the mass range $10^{12-13}$ GeV, 
where the uncertainty in $m_{\nu_{R}}$ is mainly related to the uncertainties in the neutrino masses and mixings \cite{deSalas:2017kay} through seesaw mechanism.
Third, the splitting soft mass spectrum, which is actually independent of GUT-breaking sector, 
is chosen in order to avoid the constraint from proton decay \cite{Murayama}.
Finally, identifying the whole RG trajectory between $m_Z$ and $M_{\text{U}}$  
fixes the magnitudes of $\epsilon$ terms at the scale $M_{\text{U}}$ and further $\delta y_{\alpha}$ in terms of RG running. 
The uncertainties in $\delta y_{\alpha}$ in this model 
arise from the uncertainty in the explicit value of $m_{\nu_{R}}$ and the threshold corrections at various intermediate mass scales.
For instance, the threshold correction to $\delta y_{\alpha}$ between two intermediate scales $\mu_{i}$ and $\mu_{i+1}$ is obtained by integrating the RGEs, 
which is of form $\mathcal{C}y_{\alpha}\log(\mu_{i+1}/\mu_{i})/(16\pi^{2})$, with $\mathcal{C}$ a function of gauge or Yukawa couplings.

Let us turn to the numerical calculation of $\delta y_{\alpha}$.
The \emph{left} plot in Fig.\ref{measure} shows the values of $\delta y_{\alpha}$
in the case of type-II correction in Eq.(\ref{coherence2}), 
with $r=(\upsilon^{10}_{u}/\upsilon_{u})\cdot(\upsilon^{10}_{d}/\upsilon_{d})^{-1}=8.73$ \cite{1805.10631}.
Since no public code is available for this benchmark model, 
similar to ref.\cite{1805.10631} we did the numerical analysis based on the one-loop RGEs of the SM \cite{RGESM1,RGESM2}, 
the split supersymmetry and the MSSM \cite{RGEMSSM}.
In this plot, it is clear that the parameter ranges $\mid\epsilon\mid\geq 0.2$ and $0.1\leq\mid\epsilon\mid\leq 0.2$ 
can be tested through $\delta y_{b}$ and $\delta y_{\tau}$ by the LHC with luminosity 300 fb$^{-1}$ and 3000 fb$^{-1}$, respectively.
Compared to $y_{b}$ and $y_{\tau}$, 
$y_t$ receives smallest correction but faces largest experimental uncertainties \cite{1310.8361,1902.10229}.
According to the estimates on the uncertainties above,
we expect that the theoretical uncertainty to $\delta y_{\alpha}$ is at most of order $\sim 1-3\%$.

We perform similar analysis in the \emph{right} plot in Fig.\ref{measure},
which shows the values of $\delta y_{\alpha}$ in the case of type-III correction in Eq.(\ref{coherence2}),
with $s=(\upsilon^{\overline{126}}_{u}/\upsilon^{\overline{126}}_{d})\cdot(\upsilon^{10}_{d}/\upsilon^{10}_{u})$ \cite{1805.10631}.
Compared to type II, where $\epsilon_t$ is the largest input value due to the enhancement factor $r$,
$\epsilon_{\tau}$ is the largest input value in the case of type III.
In this case, one expects larger value of $\delta y_{\tau}$,
which indicates that the same LHC limits can reach smaller parameter region $\mid\epsilon\mid \sim 0.01-0.02$, as shown in the figure.
The parameter region $\mid\epsilon\mid \geq 0.02$ can be fully covered by the LHC limit with luminosity 300 fb$^{-1}$.
Whenever the corrections to $\delta y_{\alpha}$ are roughly of same order, 
the magnitude of $\delta y_{\tau}$ at the scale $m_Z$ is always the largest.

Apart from modifying $y_{\alpha}$, the $\epsilon$-corrections also contribute to off-diagonal elements of $y_{u,d}$ 
that lead to flavor violation.
They appear even though $\epsilon^{ij}$ is diagonal at the scale $M_{\rm{U}}$ because of RGE effects \cite{1703.05873}.
In the context of type-II two Higgs doublets as we study here, 
the most stringent constraint in the parameter region with moderate or large $\tan\beta$ arises from $\rm{BR}(B_{s,d}\rightarrow \mu^{+}\mu^{-})$.
A partial reason for it is that they are enhanced by the factor $\tan\beta$,
unlike in the other cases such as $\text{Br}(t\rightarrow u_{i}h)$ ($u_{i}=\{u,c\}$) that are suppressed by $\cos^{2}(\alpha-\beta)$.
Because of the feature above, $\rm{BR}(B_{s,d}\rightarrow \mu^{+}\mu^{-})$ is actually more sensitive to parameters $\tan\beta$ and the neutral Higgs boson mass rather than the deviations of a few percent level in the Yukawa couplings in Fig.\ref{measure}.
Typically, the $\epsilon$ corrections only yield a deviation of order $\sim 0.13\%$ relative to the SM prediction $\rm{BR}(B_{s}\rightarrow \mu^{+}\mu^{-})_{\rm{SM}}=3.26 \times 10^{-9}$ for $\tan\beta=10$ and $m_{A}= m_{H}=1$ TeV, which is consistent with the LHCb limits  $\rm{BR}(B_{s}\rightarrow \mu^{+}\mu^{-})_{\text{exp}}=(2.8^{+0.7}_{-0.6})\times 10^{-9}$ \cite{1411.4413,1703.05747}.

\section{Conclusions}
Unification is an important theoretical idea of new physics beyond SM. 
Yet, there are limited ways in testing it except proton decay experiment in the last a few decades.
Unfortunately, the advance along this line is delayed due to the experimental status.
In this work, we have proposed a novel approach of probing unification through precision measurements on the Higgs Yukawa couplings especially of the third generation.

This proposal is supported by three observations.
The first observation is the appearance of unsuppressed effective operators through integrating out the heavy Higgs freedom 
$45$ or $126$ in conventional $\rm{SU}(5)$ or $\rm{SO}(10)$ models with the minimal Yukawa sector, respectively.  
The second observation is that the corrections to $y_{\alpha}$ at the scale $M_{\text{U}}$ are in three specific patterns,
as a result of either $\rm{SU}(5)$ or $\rm{SO}(10)$ embedding.
Lastly, the deviations to $y_{\alpha}$ at the scale $m_Z$ can be verified by the future LHC limits (see Fig.\ref{measure}),
although there are subject to certain uncertainties in the RG trajectory between the scales $m_Z$ and $M_{\text{U}}$.

Our analysis shows that  as a result of coherences, 
a large deviation in $y_{\tau}$ but small in $y_t$ and $y_b$
can serve as a hint of conventional realistic models of unification with the minimal Yukawa sector.
\\

\begin{acknowledgements} 
The author thanks Lisa Randall for useful discussions.
This research is supported in part by the National Natural Science Foundation of China under Grant No.11775039,
the Chinese Scholarship Council,
and the Fundamental Research Funds for the Central Universities at Chongqing University under Grant No.cqu2017hbrc1B05.
\end{acknowledgements}

\end{document}